# Modeling Language Variability

Hans Grönniger, Bernhard Rumpe

Software Engineering
RWTH Aachen University, Germany
http://www.se-rwth.de

**Abstract.** A systematic way of defining variants of a modeling language is useful for adopting the language to domain or project specific needs. Variants can be obtained by adopting the syntax or semantics of the language. In this paper, we take a formal approach to define modeling language variability and show how this helps to reason about language variants, models, and their semantics formally. We introduce the notion of semantic language refinement meaning that one semantics variant is implied by another.

## 1 Introduction

It has often been stressed that software is one of the most important drivers for innovation in many branches of industry. Developers are faced with the challenge to produce high quality, increasingly complex solutions in a short period of time.

Model-based software development is regarded as one instrument to cope with the challenges. Standard modeling languages like UML [OMG09] or domain specific languages (DLSs) are employed to increase the level of abstraction and automation while at the same time lowering the complexity. Especially in the context of robust, reliable systems development, the modeling languages used have to be defined precisely to allow for rigorous analysis of models and correct code generation.

The precise definition of a modeling language involves syntax and semantics [HR04]. Formal semantics is advantageous because it helps to avoid misunderstandings between people and may enable interoperability between tools. But even if a formal modeling language exists, a new class of systems like highly robust and reliable systems or a specific application domain may require adaptation of the language. A language may be changed to incorporate new language constructs, to disallow others for methodological or safety reasons, or to be semantically adjusted to a specific platform. This variability of a modeling language is subject of the paper.

We provide a formal account on language variability based on our classification in [CGR09]. On the one hand, the formalization brings light into how a language can be adopted to specific requirements. On the other hand, it servers as a basis to define language variants formally. This allows us to reason about language (especially semantic) variants.

The paper is structured as follows. The basic constituents (syntax, semantics) of a modeling language that may be subject to variability are introduced in Section 2. In Section 3, a formal characterization of language variants and a method to define variants is presented. As an example application, we outline how semantic variants can be compared formally in Section 4. In Section 5, we discuss related work. Section 6 concludes the paper.

## 2 Language Constituents

A precise definition of a modeling language consists of the following elements, see also [HR04,CGR09].

*Concrete Syntax* The concrete syntax is the representation of the model with which a user interacts. This may be a graphical or textual notation or a mixture of both. We denote the set of all models of a modeling language in concrete syntax by $\mathcal{CS}$.

*Abstract Syntax* The abstract syntax represents the structural essence of a language [Wil97]. For a textual syntax this may be given as abstract syntax trees generated by a parser. In case of graphical models, metamodels (e.g., defined in MOF [OMG06]) are typically used. The set of all models of a modeling language in abstract syntax is denoted by $\mathcal{AS}$.

Additionally, a set of well-formedness rules or context conditions are defined to rule out certain models based on syntactic criteria. We assume a predicate

$$\text{wellformed} : \mathcal{AS} \to bool$$

to decide if a model is well-formed. The set of all well-formed models $\mathcal{AS}^{\text{wf}}$ of a language hence is

$$\mathcal{AS}^{\text{wf}} = \{m \in \mathcal{AS} \mid \text{wellformed}\, m\}$$

A model in concrete syntax is associated with (or mapped to) a model in abstract syntax. Since typically not all models from $\mathcal{CS}$ are well-formed, we obtain a partial mapping from concrete to abstract syntax:

$$\text{p} : \mathcal{CS} \rightharpoonup \mathcal{AS}^{\text{wf}}$$

*Reduced Abstract Syntax* It is often advisable to reduce the number of language constructs by further constraining the set $\mathcal{AS}^{\text{wf}}$ to a subset $\mathcal{AS}^{\text{red}} \subseteq \mathcal{AS}^{\text{wf}}$. For some models in $\mathcal{AS}^{\text{wf}}$ there will be a syntactic transformation t to convert it into the reduced abstract syntax, i.e.,

$$\text{t} : \mathcal{AS}^{\text{wf}} \rightharpoonup \mathcal{AS}^{\text{red}} \text{ with } \mathcal{AS}^{\text{wf}} \supseteq \text{dom}(\text{t}) \supseteq \mathcal{AS}^{\text{red}}$$

The reduction of the abstract syntax might be useful for several reasons. One is that is eases semantics definition. A more detailed explanation will be given in the next section.

*Semantic Domain* By mapping models to elements of a semantic domain $\mathcal{S}$, the models obtain their meaning. The semantic domain is required to be well-known and understood and it should be based on a well-defined mathematical theory.

Our approach to semantics uses the system model [BCGR09a,BCGR09b] which characterizes the structure, behavior, and interaction of objects in object-based systems. The definitions are built on simple mathematical concepts like sets, relations, and functions. It is important to note that one element in the system model represents a single, complete object-based system. This means that the meaning of a model is directly represented as properties of possible implementations. The system model is underspecified to allow, for example, freedom of implementation when mapping a model to executable code.

*Semantic Mapping* The semantic mapping sem finally relates models of the reduced abstract syntax to elements of the semantic domain. Characteristic of our loose approach is a set-valued or predicative semantic mapping of the form

$$\text{sem} : \mathcal{AS}^{\text{red}} \to \wp(\mathcal{S})$$

$\wp(X)$ denotes the set of all subsets of $X$ (power set). The semantics of a model $m$ is therefore the set $\text{sem}(m)$ of elements in the domain $\mathcal{S}$. If the system model is used for $S$, then the model's meaning is the set of all possible realizations.

Using the system model as a single semantic domain and the set-valued semantic mapping enable a straightforward treatment of composition and refinement of possibly incomplete and underspecified models of various modeling languages [Rum96]. For example, the integrated semantics of models $m_1, \ldots, m_n$ from possibly different languages is given as $\text{sem}_1(m_1) \cap \ldots \cap \text{sem}_n(m_n)$. In the same way, a model $m'$ is a refinement of model $m$, exactly if $\text{sem}(m') \subseteq \text{sem}(m)$.

## 3 Language Variants

A modeling language should be defined precisely but should not be completely fixed. Sustaining a certain degree of flexibility regarding a language's syntax or semantics allows for adapting it to project or domain specific needs, or to enable modeling of new classes of systems. This idea has also been incorporated in the definition of UML where the informal semantics is equipped with semantic variation points subject to specific interpretation. We take a formal approach to define the possible variability in a language definition thereby substantiating our classification in [CGR09]. Afterwards, we present an intuitive way to document language variants.

### 3.1 Classification of Language Variability

In the previous section, we defined the constituents of a modeling language and their relations. To summarize, we have the sequence

$$\mathcal{CS} \stackrel{\text{p}}{\rightharpoonup} \mathcal{AS}^{\text{wf}} \stackrel{\text{t}}{\rightharpoonup} \mathcal{AS}^{\text{red}} \stackrel{\text{sem}}{\to} \wp(S)$$

In this section, we discuss means to define variants of a modeling language by adopting one or more elements of the above sequence.

*Presentation Variability* A modeling language may offer *presentation options*, a term also coined in the UML standard. Presentation options allow for representing models differently in concrete syntax without changing a model's abstract syntax. Formally, a language contains presentation options, if

$$\exists m_1, m_2 \in \mathcal{CS} : m_1 \neq m_2 \land \mathrm{p}(m_1) = \mathrm{p}(m_2)$$

For example, we have different ways to represent a public class modifier in UML: We can use the keyword `public` but equivalently the symbol `+`. Variants of presentation options result in changes of $\mathcal{CS}$ and p, say $\mathcal{CS}_v$ and $\mathrm{p}_v$, by introducing, eliminating or changing existing ones. Models contained in both variants still have the same abstract syntax:

$$\forall m \in \mathcal{CS}_v \cap \mathcal{CS} : \mathrm{p}_v(m) = \mathrm{p}(m)$$

Additionally, every model can be expressed without choosing the presentation option variant:

$$\forall m_1 \in \mathrm{dom}(\mathrm{p}_v) : \exists m_2 \in \mathrm{dom}(\mathrm{p}) : \mathrm{p}_v(m_1) = \mathrm{p}(m_2)$$

Another form of presentation variability is what we call *abbreviations* or *extended constructs*: The syntax may contain certain constructs that help to enhance readability and comfort but which can be eliminated by some syntactic transformation t without loosing expressiveness of the language. All models which do not use extended constructs remain identical under t, i.e.,

$$\forall m \in \mathcal{AS}^{\mathrm{red}} : t(m) = m$$

The models that actually get transformed are contained in $\mathrm{dom}(\mathrm{t}) \backslash \mathcal{AS}^{\mathrm{red}}$. Variability in abbreviations means adapting $\mathcal{AS}^{\mathrm{wf}}$ and t, to $\mathcal{AS}^{\mathrm{wf}}_v$ and $\mathrm{t}_v$ say. Consider, for example, a reduced abstract syntax for Statecharts $\mathcal{AS}^{\mathrm{red}}$ which contains flat automata only. Hierarchy can be added to or removed from Statecharts without changing expressiveness [Rum04], but we obtain a larger set of expressible models when adding hierarchy, i.e., $\mathcal{AS}^{\mathrm{wf}}_v \supseteq \mathcal{AS}^{\mathrm{red}}$. Models that do not contain an extended construct variant (e.g., hierarchy) are transformed equally under $\mathrm{t}_v$:

$$\forall m \in \mathrm{dom}(\mathrm{t}_v) \cap \mathrm{dom}(\mathrm{t}) : \mathrm{t}_v(m) = \mathrm{t}(m)$$

And we can still represent each model without the abbreviation:

$$\forall m_1 \in \mathrm{dom}(\mathrm{t}_v) : \exists m_2 \in \mathrm{dom}(\mathrm{t}) : \mathrm{t}_v(m_1) = \mathrm{t}(m_2)$$

As abbreviations do not show up in the reduced abstract syntax, semantics of theses constructs is defined in two-step, the first one being the transformation to $\mathcal{AS}^{\mathrm{red}}$ for which semantics is defined via the semantic mapping sem. Summarizing, variants of presentation options have an effect on the concrete syntax. Variants of abbreviations have an effect on the full abstract syntax. Both do not change the reduced abstract syntax and are called presentation variability.

*Syntactic Variability* We now consider language variants that also have an impact on the reduced abstract syntax $\mathcal{AS}^{\text{red}}$. The syntax of a language may allow the use of *stereotypes*. A set of defined stereotypes (e.g., as part of a profile in case of UML) is a syntactic variant of the language. We assume a function variant allowedStereotypes$_v$ that checks if only the chosen stereotypes are used, i.e.,

$$\mathcal{AS}^{\text{red}}_v = \{m \in \mathcal{AS}^{\text{red}} | \text{allowedStereotypes}_v(m)\}$$

Another form of syntactic variability is given by so called *language parameters*, also termed language embedding in [KRV08]. Consider, for example, the language of Statecharts in which transitions may be guarded by a precondition. The language in which this condition is expressed is not specified. Hence, a syntax can be equipped with parameters $\mathcal{AS}^{\text{red}}(p_1, \ldots, p_n)$. Variants can then be specified by assigning concrete languages to the parameters $p_1, \ldots, p_n$.

As a last form of syntactic variability, we consider general *language constraints*. A language is further constrained to disallow certain models syntactically. It may be the case that this results is a less expressive language. Formally, a variant $\mathcal{AS}^{\text{red}}_v$ is given by models which fulfill further constraints stated, for example, in the predicate constr$_v$:

$$\mathcal{AS}^{\text{red}}_v = \{m \in \mathcal{AS}^{\text{red}} | \text{constr}_v(m)\}$$

The expressiveness of the language is preserved if

$$\forall m_1 \in \mathcal{AS}^{\text{red}}_v : \exists m_2 \in \mathcal{AS}^{\text{red}} : \text{sem}(m_1) = \text{sem}(m_2)$$

It is, for example, the goal of modeling or programming guidelines [Mat07,MIS] to restrict the use of certain (e.g., unsafe) language constructs while preserving the expressiveness. Restricting the expressiveness might be useful in situations in which a target platform may not be powerful enough to implement the models.

*Semantic Variability* While UML only uses the term semantic variation point, we further subdivide semantic variability into *semantic mapping variability* and *semantic domain variability*. A helpful analogy might be to see the variability of the semantic mapping similar to configuration options of a code generator while variability of the semantic domain has its analogy with properties of an underlying run-time system or target platform.

By selecting variants for the semantic domain $\mathcal{S}$, we obtain an adapted domain $\mathcal{S}_v$ in which elements have certain additional properties, for example, encoded in a predicate prop$_v$:

$$\mathcal{S}_v = \{s \in \mathcal{S} | \text{prop}_v(s)\}$$

Regarding semantic domain variability, the system model already contains explicit variability in form of extensions through optional definitions. It provides, for instance, different notions of type-safe method overriding or optional constraints to allow single inheritance only.

Variants of a semantic mapping arise as alternative definitions of (parts of) the semantic mapping, for example

$$\text{sem}_{v1}, \text{sem}_{v2} : \mathcal{AS}^{\text{red}} \to \wp(\mathcal{S})$$

Considering a Statecharts semantics again, a mapping variant could be the different choices of representing Statecharts states (syntax) as, for example, a simple enumeration in a class or using the state pattern [GHJV95].

Note that semantic variability is transparent to the modeler. But it may be necessary to allow the modeler to select one or the other interpretation of a construct. We propose to model these interpretation choices as syntactic variability by providing corresponding stereotypes. A modeler can then select the semantics of certain constructs by using appropriate stereotypes. With this approach, we transfer semantic variation points to syntactic ones.

### 3.2 Documentation of Language Variability

We propose to model variation points and variants in a language by feature diagrams [CE00]. Fig. 1 contains a feature diagram representing a generic structure to model variants of a language L. We do not show concrete variants which depend on a specific language and which would be inserted under the corresponding nodes.

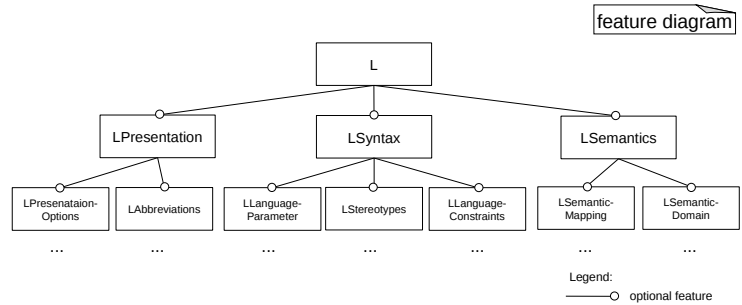

**Fig. 1.** Template to document variability of a language L

A supplement description of the variability can be given to explain their raison d'être and to point to formal definitions of the variants or other documentation.

## 4 Comparison of Semantic Variants

Using our formal notion of language variants, it is possible to compare language variants formally and derive properties of the (relationship between) variants.

Consider two semantic variants of the same language, e.g.,

$$\mathrm{sem}_{v1} : \mathcal{AS}^{\mathrm{red}} \to \wp(\mathcal{S}_{v1})$$
$$\mathrm{sem}_{v2} : \mathcal{AS}^{\mathrm{red}} \to \wp(\mathcal{S}_{v2})$$

An interesting property is if variant $v2$ is a *semantic language refinement* of the semantic variant $v1$. Note that we discuss language refinement here and do not talk about refinement of models or the modeled system.

We define that language variant $v2$ is a semantic language refinement of variant $v1$ exactly if for all models the sets generated by the respective semantic mapping are in a subset relation, i.e.,

$$\forall m \in \mathcal{AS}^{\mathrm{red}} : \mathrm{sem}_{v1}(m) \supseteq \mathrm{sem}_{v2}(m)$$

This implies that all properties $\phi$ of a model $m$ which hold in variant $v1$ are preserved in variant $v2$:

$$\forall s \in \mathrm{sem}_{v1}(m) : \phi(s) \implies \forall s \in \mathrm{sem}_{v2}(m) : \phi(s)$$

Semantic language refinement is an important property if we consider for example tool integration. Assume that one tool for formal analysis uses (and correctly implements) language variant $v1$. Another tool for code generation correctly implements variant $v2$. If we show that variant $v1$ is a language refinement of $v2$ then we can be sure that analysis results obtained by the first tool are preserved in the second tool.

## 5 Related Work

Presentation and semantic variants are also covered informally in the UML standard [OMG09]. We state precisely what kinds of variability may be found in a modeling language and document variants using feature diagrams. Feature diagrams are also used in [Völ08] to define a family of architecture description languages. Formal semantics is not addressed. In the area of semantics, semantic variability is covered to some extent. Template semantics [NAD03] as well as templatable metamodels [CMTG07] can be used to describe semantics with variation points. Non of the mentioned work discusses the possibility to compare language variants. [TA06] examines different variants of formal Statecharts semantics but does not address formal relationships between the variants. Informal comparisons of Statecharts variants can, for example, be found in [Bee94,CD07].

## 6 Conclusion

We have formally described the constituents of a modeling language and how they can be varied to obtain modeling language variants. As an example application of precise modeling language variants, we have introduced the notion of semantic language refinement. Given two semantics variants of a language this notion defines if it is safe to use the one instead of the other variant. Future work is concerned with investigating other relationships between language variants. Additionally, this work needs to be applied to, for example, the UML, or to various domain specific languages and needs to be brought into practice by appropriate tool support.